\documentstyle[12pt]{article}

\newcommand{\newsection}{    % Numeration of eqs. is automatic
\setcounter{equation}{0}
\section}
\newcommand{\tr}[1]{\,{\rm tr}\,#1}
\newcommand{\ntr}[1]{\,\frac {\rm tr}{N}\,#1}
\def\e{{\,\rm e}\,}

\def\be{\begin{equation}}
\def\ee{\end{equation}}
\def\bea{\begin{eqnarray}}
\def\eea{\end{eqnarray}}
\def\LA{\left\langle}
\def\RA{\right\rangle}

\newcommand{\rf}[1]{(\ref{#1})}
\newcommand{\eq}[1]{Eq.~(\ref{#1})}

\def\a{\alpha}
\def\b{\beta}
\def\L{\Lambda}
\def\l{\lambda}

\def\om{\omega}
\newcommand{\ie}{{\it i.e.}\ }

\newcommand{\p}{{\prime}}
\newcommand{\ra}{\rightarrow}
\def\appendix#1{
\addtocounter{section}{1}
\setcounter{equation}{0}
\renewcommand{\thesection}{\Alph{section}}
\section*{Appendix \thesection\protect\indent #1}
\addcontentsline{toc}{section}{Appendix \thesection\ \ \ #1}
}
\newcommand{\ci}{\int_{C_1}\frac{d\omega}{2\pi i}}
\newcommand{\cii}{\int_{Cv_2}\frac{d\omega}{2\pi i}}

\newcommand{\non}{\nonumber \\*}
\let\la=\lambda
\let\La=\Lambda
\newcommand{\re}[1]{(\ref{#1})}
\newcommand{\real}{{\cal R}}

\begin{document}

\begin{titlepage}
\begin{flushright}
UBCTP-xxx-93 \\ITEP-YM-3-93 \\ April, 1993
\end{flushright}
\vspace{1cm}

\begin{center}
{\LARGE Correlators of the Kazakov--Migdal Model} \end{center} \vspace{.5cm}
\begin{center}
{\large M.I.~Dobroliubov}$~^{a,}
$\footnote{Permanent address: Institute
for Nuclear Research, Academy of Sciences of Russia,
Moscow, Russian Federation}$^{,2}$, {\large Yu.~Makeenko}$~^b$,
{\large G.W.~Semenoff}$~^{a,}$\footnote{This work was supported in part by the
Natural Sciences and Engineering Research Council of
Canada}
\end{center}
\begin{center}
{}$^a$~{\it Department of Physics, University of British Columbia\\
Vancouver, British Columbia, V6T 1Z1 Canada} \\ \mbox{} \\ {}$^b$~{\it
Institute of Theoretical and Experimental Physics} \\ {\it
117259 Moscow, Russian Federation}
\end{center}

\vskip 1.5 truein
\begin{abstract}
\noindent
We derive loop equations
for the one-link correlators of gauge and scalar fields in the
Kazakov--Migdal model.  These equations determine the solution of the
model in the large $N$ limit and are similar to analogous equations
for the Hermitean two-matrix model.  We give an explicit solution of
the equations for the case of a Gaussian, quadratic potential.  We
also show how similar calculations in a non-Gaussian case reduce to
purely algebraic equations.
\end{abstract}
\end{titlepage}

\newpage
\section{Introduction}
Solving QCD in the limit of a large number of colors is a classic
problem \cite{lgn}.  Recently, Kazakov and Migdal~\cite{KM92} have
noted that it may indeed be possible to find an analytic solution of
the large $N$ limit of a certain version of induced $SU(N)$ lattice
gauge theory. In their model, the Yang-Mills action is absent at
lattice distance scales and is induced in the continuum limit by the
interaction of the gauge field with a heavy Hermitean matrix-valued
scalar field.  The eigenvalues of the scalar field behave as the
classical ``master field'' in the large $N$ limit.  The possibility of
solving this model has attracted wide attention and the task of
unraveling its physical content is a subject of ongoing research.
Besides being a candidate for QCD, it is an interesting
example of a matrix model in dimensions greater than one for which a
solution in the large N limit may be attainable.

The Kazakov-Migdal model has action
\be
Z=\int \prod_xd\phi_x\prod_{\!\! xy}[ dU_{xy}] e^{S[\phi,U]} \, ,
\label{partition}
\ee
\noindent
where
\be
 S[\phi,U]=-\sum_xN{\rm tr}V(\phi_x) +\sum_{\!\! xy}N{\rm
tr}\phi_xU_{xy}\phi_yU^{\dagger}_{xy}~~,
\label{action}
\ee
$\phi_x$ and $U_{xy}$ ($=U^{\dagger}_{yx}$) are $N\times N$ Hermitean
and unitary matrices, which live on lattice sites $x$ and links ${xy}$
between neighboring sites, respectively.  Here $[dU]$ denotes the Haar
measure for integration over $U(N)$.  Using the Itzykson-Zuber
integral \cite{itzub}
\be
I[\phi,\chi]\equiv\int[dU]e^{N{\rm tr}\phi U\chi U^{\dagger}}
=\frac{\det^{ij}e^{N\phi^i\chi^j}}{\Delta(\phi)\Delta(\chi)} \, ,
\ee
\noindent
where $\phi^i,\chi^j$ are the eigenvalues of $\phi$ and $\chi$ and
$\Delta(\phi)=\prod_{i<j}(\phi^i-\phi^j)$ is the Vandermonde
determinant, (\ref{partition}) can be presented as the effective
theory for a scalar field
\be
Z=\int
\prod_xd\phi^i_xe^{S_{\rm eff}[\phi]}
\ee
with the action
\be
S_{\rm eff}[\phi]= -N\sum_{i,x}V(\phi^i_x)
+\sum_x \ln\Delta^2(\phi_x)+\sum_{\!\! xy} \ln I[\phi_x,\phi_y]
\ee
(The Vandermonde determinants result from converting the integration
measure for matrices to one for eigenvalues.)

When $N$ is large this integral can be evaluated in the saddle-point
approximation.  The saddle-point equation is
\be
\frac{2D}{N}\frac{\partial}{\partial\phi^i}\ln I[\phi,\chi]\vert_{\chi=\phi}
=V'(\phi^i)-\frac 2N \sum_{j\neq i}\frac{1}{\phi^i-\phi^j}
\label{saddle point}
\ee

This model was formally solved by Migdal \cite{Mig92a}, who derived a
singular nonlinear integral equation for the spectral density of the
eigenvalues.  This equation is rather complicated and its explicit
solution even for the simplest Gaussian case proved to be
non-trivial~\cite{Gro92}.  One of the authors has
proposed~\cite{Mak92} an approach to this problem which employs the
method of loop equations. In this paper we shall generalize these
ideas to derive a loop equation for arbitrary one-link correlation
functions.  Its solution provides the spectral density of eigenvalues
of $\phi$ as well as the full set of one-link correlators of the gauge
and scalar fields.  We shall obtain an explicit solution of these
equations in the case of a quadratic potential and reduce the more
complicated case of a general potential to the solution of algebraic
equations.

The importance of the correlators of a pair of the gauge fields was
discussed in \cite{KMSW92a} where it was argued that, to a large
extent, they determine the continuum limit of the theory.  More
recently, in \cite{DKSW93} it was shown that they could also play an
essential role in resolving the problem of $Z_N$ gauge invariance of
the lattice model (\ref{partition})~\cite{KSW92,KhM92}.
Consider the correlator (our
normalization differs from the original normalization of
Ref.~\cite{KMSW92a} by an extra $1/N$)
\be
{1\over N} C^{ij}[x,y]\delta_{il}
\delta_{jk}=\LA U^{ij}_{xy}U^{kl}_{yx}\RA \equiv\frac{\int d\phi [dU]\e^S
 U^{ij}_{xy}U^{kl}_{yx}}
{\int
d\phi[dU] \e^S}\,.
\label{Cij}
\ee
Here, the delta-functions on the right-hand side arise from the gauge
invariance of the integral. It also follows from gauge invariance that
$C_{ij}$ is symmetric.  Unitarity of $U$ implies the sum rule
\be
\frac{1}{N} \sum_i
C_{ij}=1~~.
\label{sum rule}
\ee

In the large $N$ limit, the integration over the scalar field is
performed by replacing it with its saddle point value (this should
properly be done after $\phi$ is diagonalized).  Then, the expectation
value of a quantity such as $UU^{\dagger}$ factorize into one-link
expectation values,
\be
\frac{1}{N}
 C_{ij}=\frac { \int [dU]\e^{N\tr \phi_xU\phi_yU^{\dagger}} \vert
U^{ij}\vert^2 } {\int [dU]\e^{N\tr \phi_x U\phi_y U^{\dagger}} } \, .
\ee
{}From this equation we can derive the quantity
\be
\frac{1}{N}\sum_j C_{ij}\phi_j = \frac{1}{N}
\frac{\partial}{\partial\chi_i}\ln I[\phi,\chi]\vert_{\chi=\phi} \, ,
\ee
which can be obtained from the saddle-point equation \re{saddle point} as
\be
\frac{1}{N} \sum_j C_{ij}\phi_j =\frac{1}{2D}\left( V'(\phi_i)- \frac{2}{N}
\sum_{j\neq i} \frac{1}{(\phi_i-\phi_j)}\right) \, .
\label{mfe}
\ee

Formal expressions for the integrals \re{Cij} at finite $N$, in terms of the
eigenvalues $\phi_i$  have been
found in \cite{MorShat}.  Those
expressions  are
potentially very useful when $N$ is small rather than
large.  Here, to analyze the infinite $N$ limit, we shall take a
different approach which requires a simultaneous solution for the
correlator and the eigenvalue distribution.

In the large $N$ limit, it is necessary to replace the index of the
eigenvalues of $\phi$ with a continuous label.  This is done by
introducing the eigenvalue density, $\rho(\alpha)$ such that
$\rho(\alpha)d\alpha$ is the number of eigenvalues in the interval
$[\alpha,\alpha+d\alpha]$.  The spectral density typically has support
in a finite interval, $[a,b]$, and is normalized so that
\be
\int_a^b d\alpha \rho(\alpha) =1~~,~~~~
\endequation
Also, the normalization of the correlator \re{sum rule} becomes
\equation
\int_a^b d\alpha\rho(\alpha)C(\alpha,\beta)=1\,.
\ee
The saddle-point equation \re{mfe} is
\be
\int_a^b d\beta\rho(\beta)C(\alpha,\beta)\beta=\frac{1}{2D}
\left(V'(\a)- 2\real E_\a \right)
\label{mfe-c}
\ee
where $\real E_\a$ is the real (and continuous across the support of
$\rho$) part of the analytic function (of the complex variable
$\lambda$)
\be
E_\l\equiv \LA\frac{\rm tr}{N}\frac{1}{\l-\phi_x}\RA~=~\int_a^b
d\alpha\rho(\alpha)\frac{1}{\l-\alpha} \,.
\label{defE}
\ee

Equations of motion (\ref{mfe}), (\ref{mfe-c}) imply that,
anywhere that the single-link operator
$U_{xy}\phi_yU_{yx}$ appears with other operators inside a gauge
invariant correlation function, and $U_{xy}$ and $\phi_y$ appear
no-where else in the other operators in the correlators, in the
saddle-point approximation we can replace the {\it diagonal}\/ elements of
the operator
$U_{xy}\phi_yU_{yx}$ by the operator
\be
F(\phi_x) \equiv \frac{1}{2D}\left(
V'(\phi_x)-\tilde{V}'(\phi_x)\right) ~,
\label{Lambda}
\ee
where $\tilde{V}'(x)$ is the analytic continuation of $2 \real E_\la$
from the cut on the whole complex plane, so that on the cut
\be
\tilde{V}'(\a) = 2\real E_\a ~~.
\label{tildeV}
\ee
This looks like the saddle point equation for the eigenvalue
distribution of a
one-matrix model with potential $\tilde V(\phi)$ (see also \eq{EE}).

\newsection{Loop equations for one-link correlators}

\subsection{The two analytic functions}

The central quantities of interest in this Section are the two
analytic functions: $E_\l$, which is defined by \eq{defE}, and
\be
G_{\nu\lambda}\equiv\LA\frac{\rm
tr}{N}\frac{1}{\nu-\phi_x}U_{xy}\frac{1}{\lambda-\phi_y}U_{yx}\RA~ =~
\int_a^b d\alpha \rho(\alpha)\int_a^b d\beta
\rho(\beta)\frac{C(\alpha,\beta)}{(\nu-\alpha)(\lambda-\beta)}.
\label{defG}
\ee
It is important to note that, due to invariance of the group measure
under transformations $U\ra U^\dagger$, $G_{\nu\la}$ is symmetric in
$\nu$ and $\la$.

These two functions are analytic on the whole complex plane excluding
cut singularities in the real axis in interval $[a,b]$ (the support of
$\rho$). At the cut they have following discontinuities
\be
 E_{\a\pm i0} = {\cal P} \int_a^b d\beta\rho(\beta) \frac{1}{\a-\beta}
\mp i\pi\rho (\a)~,
\label{discE}
\ee
and
\be
G_{\a \pm i0, \la} = {\cal P} \int_a^b d\beta \rho (\beta)
\frac{{\cal G}_\la(\beta)}{\a-\beta} \mp
i\pi\rho (\a) {\cal G}_\la(\a) ~,
\ee
where
\bea
 {\cal G}_\la (\a) \equiv \int_a^b d\!\b \rho (\b) \frac{C(\a,\b)}{\la-\b}
\label{discG}
\eea
Note that this defines a function of the matrix $\phi_x$, which we can
also view as a 1-link expectation value
\equation
{\cal G}_\la(\phi_x)=\langle U_{xy}\frac{1}{\lambda-\phi_y}U_{yx}\rangle_{1L}
\endequation
(where the bracket denotes integration over $U_{xy}$ and $\phi_y$ is
evaluated at the saddle point).  The function ${\cal G}_\lambda(\phi)$
was introduced by Migdal \cite{Mig92a} and was the key component of
his approach (see Sect.~\ref{RG} below).

Since the support of $\rho$ is finite, the functions $E_\nu$ and
$G_{\nu\la}$ have the following asymptotic expansions
\be
E_\nu=\frac 1\nu + \sum_{k=1}^\infty \frac{E_{k}}{\nu^{k+1}}~~,~~~~~~~
E_k=\LA\ntr \phi^k \RA
\label{bc1}
\ee
and
\be
G_{\nu\l}=\frac{E_\l}{\nu} + \sum_{n=1}^\infty
\frac{G_n(\l)}{\nu^{n+1}} ~~,~~~~~~~
G_n(\l)=\LA\ntr{}\Big( \phi^n U\frac{1}{\l-\phi} U^\dagger \Big)\RA
\label{bcG}
\ee
As we see from \eq{bcG}, asymptotics of $G_{\nu\la}$ completely fixes
$E_\la$.

\subsection{Loop equation for $G_{\nu\lambda}$}

Our main equation is derived from the identity
\begin{eqnarray}
& \frac{N^{-2}}{Z}\int d\phi
[dU]{\rm tr}\frac{\partial}{\partial\phi^{ij}_x} \e^S\left(
\frac{1}{\nu-\phi_x}U_{xy}\frac{1}{\lambda-\phi_y}U_{yx}\right)^{ij}
=\LA\frac{\rm tr}{N}\frac{1}{\nu-\phi_x}\frac{\rm
tr}{N}\frac{1}{\nu-\phi_x}U_{xy}\frac{1}{\lambda-\phi_y}U_{yx}\RA
\nonumber\\
& -\LA\frac{\rm tr}{N}
\frac{V'(\phi_x)}{\nu-\phi_x}U_{xy}\frac{1}{\lambda-\phi_y}U_{yx}\RA
+\sum_z \LA\frac{\rm tr}{N}
U_{xz}\phi_zU_{zx}\frac{1}{\nu-\phi_x}U_{xy}\frac{1}{\lambda-\phi_y}U_{yx}\RA
{}~~.
\label{LE G}
\end{eqnarray}

The first term on the right-hand-side can be factored into two terms.
This factorization is valid in the large $N$ limit because of the
saddle point evaluation of the integral over $\phi$.  The result is a
product of two quantities
\be
\LA\frac{\rm tr}{N}\frac{1}{\nu-\phi_x}\frac{\rm
tr}{N}\frac{1}{\nu-\phi_x}U_{xy}\frac{1}{\lambda-\phi_y}U_{yx}\RA
=E_\nu G_{\nu\lambda}\,.  \ee

The third term on the right-hand-side contains link operators which
connect $x$ to all neighboring points.  One of these connects $x$ to
$y$.  It contributes
\be
\LA\frac{\rm tr}{N}
\frac{1}{\nu-\phi_x}U_{xy}\frac{\phi_y}{\lambda-\phi_y}U_{yx}\RA
=-E_\nu+\lambda G_{\nu\lambda}
\ee
There are also $2D-1$ links which connect to other sites.  For these
one can use the fact that the quantity $U_{xz}\phi_zU_{zx}$ inside the
expectation value bracket can be replaced by $F(\phi_x)$
(cf. \eq{Lambda}).  With this input we present \eq{LE G} in the form
\bea
0 = E_\nu G_{\nu \la} - \LA\ntr  L(\phi)  \frac{1}{\nu-\phi} U
\frac{1}{\la-\phi} U^\dagger\RA + \la G_{\nu \la} -
E_\nu
\eea
Where, for brevity of notation, we have defined
\be
 L(\om)\equiv V^\prime(\om)-(2D-1)F(\om) =
\tilde{V}^\prime (\om)+F(\om)\,.
\label{defL}
\ee
If we assume that $ L(\omega)$ has no singularities in the
interval $[a,b]$, this equation can be written in the integral form
\footnote{An equivalent set of equations for an Hermitean two-matrix
model was derived by Staudacher~\cite{Sta93} and by
Ambjorn and Kristjansen~\cite{AK}. The point is
that for the two-matrix model, which is associated with $D=1/2$ in the
formulas, one gets from \eq{defL} that $
L(\om)={V}^\p(\om)$.}
\be
 \int_{C_1} \frac{d \om}{2\pi i}
\frac{L(\om)}{\nu - \om}G_{\om \la}=
E_\nu G_{\nu \la} + \la G_{\nu \la} - E_\nu \, ,
\label{main}
\ee
where the contour $C_1$ encircles the cut of the function $G_{\nu\la}$
(which coincides with the support of the spectral density).

\subsection{Loop equation for $E_\nu$}

Consider the limit $\la\ra\infty$ in \eq{main}. The leading term gives
\be
G_1(\nu) = \int_{C_1} \frac{d \om}{2\pi i}
\frac{ L(\om)}{\nu - \om} E_\om - E^2_\nu \, .
\label{1}
\ee
{}From the definition \re{defG} of $G_{\nu\la}$
\bea
 G_1(\nu) = \LA \ntr \frac{1}{\nu-\phi} U\phi U^\dagger\RA =
\LA \ntr \frac{1}{\nu-\phi} F(\phi) \RA
%\non
 = \int_{C_1} \frac{d \om}{2\pi i}
\frac{F(\om)}{\nu - \om} E_\om
\label{defG1}
\eea
Plugging this equation back to \eq{1} and taking into account the definition
\re{defL} one gets the equation for $E_\nu$ first obtained in \cite{Mak93}
\be
 \int_{C_1} \frac{d \om}{2\pi i}
\frac{\tilde{V}^\prime(\om)}{\nu - \om} E_\om
= E^2_\nu \, ,
\label{EE}
\ee
which looks like the large-$N$ loop equation for an Hermitean one-matrix model
with the potential $\tilde{V}$. However, in contrast to the one-matrix model
where $\tilde{V}$ is usually a polynomial, we shall consider below the case of
non-polynomial $\tilde{V}$ which has singularities on the complex plane outside
the support of $\rho$.

\newsection{Relation to the Migdal's approach \label{RG} }

The purpose of this section is to demonstrate how, within our
approach, one can recover the previous approach to solution of this
model due to Migdal \cite{Mig92a}.  For simplicity we restrict
ourselves to the case where the density of eigenvalues, $\rho$, has
support in a single interval $[a,b]$.  It is straightforward to
generalize to the situation where the support is in two or more
intervals.

Consider our equation \re{main}. By construction (cf. \eq{defG}), the
function $G_{\nu \la}$ has singularities only on the support of the
spectral density $\rho$, where it has a cut.  The contour $C_1$
encircles this cut, and the parameters $\nu$ and $\la$ lie outside of
it. Recall as well that the function $L(\omega)$ is assumed to be
continuous across the cut (cf. \eq{defL}).

Let us compress the contour $C_1$ to make it coincide with the cut.
Then only the discontinuity of the function $G_{\omega \la}$ at this
cut (cf.
\eq{discG}) will contribute to the integral on the l.h.s. of
\eq{main}. Then \eq{main} can be rewritten in the  form
\be
\int_a^b dy \rho (y)  L(y) \frac{{\cal G}_\la(y)}{\nu-y} =
E_\nu G_{\nu \la} + \la G_{\nu \la} - E_\nu \,.
\label{new eq}
\ee

Consider the limiting case of (\ref{new eq}) when $\nu$ approaches the
cut singularity on the real axis from above:
\bea
&{\cal P} \int_a^b dy \rho (y) L(y) \frac{{\cal G}_\la(y)}{\nu-y} -
i\pi \rho(\nu)  L(\nu) {\cal G}_\la(\nu) = & \\
& (\real E_\nu - i\pi\rho(\nu) +\la) \left( {\cal P}
\int_a^b dy \rho (y) \frac{G_\la(y)}{\nu-y} -
i\pi \rho(\nu) {\cal G}_\la(\nu) \right) - \real E_\nu + i\pi\rho(\nu) \, . &
\nonumber
\label{above}
\eea
Writing a similar equation in the limit when $\nu$ approaches the cut
from below and subtracting from it \re{above}, we finally get
\be
{\cal P}
\int_a^b dy \rho (y) \frac{{\cal G}_\la(y)}{y-\nu} = -1 + \left( \la+\real
E_\nu
 -
L(\nu) \right) {\cal G}_\la(\nu) \label{fin eq}
\ee
Let us now make use of the
relation between the real part of the function $E_\nu$ and the function
$F(\nu)$
(cf. Eqs.~\rf{Lambda}, \rf{tildeV}, \rf{defL})
\be
L(\nu)- F(\nu)=
\tilde{V'}(\nu) = 2 \real E_\nu \, ,
 \label{prop}
\ee
where $F(\nu)$ is the logarithmic derivative of the Itzykson-Zuber integral
which enters the saddle-point equation~\rf{saddle point}. Plugging \re{prop}
into \re{fin eq},
we find exactly the basic equation of the Riemann-Hilbert approach of
\cite{Mig92a}
\be
{\cal P} \int_a^b dy \rho (y) \frac{{\cal G}_\la(y)}{y-\nu} =
-1 + (\la - F(\nu) - \real E_\nu ) {\cal G}_\la(\nu) \, .
\label{Migdal}
\ee

\newsection{The exact solutions}

\subsection{The general potential}

Let us consider an arbitrary potential $V(\om)$, which
is associated with a general $L(\om)$ (for simplicity we consider the case when
$L(\om)$ is analytic on the whole complex plane, except, may be, infinity)
\be
L(\omega)=\sum_{m=0}^\infty L_{m}\omega^{m} \,.
\label{def L}
\ee
The equation for $G$~\rf{main} in this case is
written explicitly
\be
\left(\lambda + E_\nu -  L(\nu) \right)G_{\nu\lambda}~=~E_\nu -R_\l(\nu)
\label{gen eq for G}
\ee
where $R_\l(\nu)$ is given by
\be
R_\l(\nu) \equiv -\cii \frac{L(\om)}{\nu-\om} G_{\om\l} =
\sum_{m=1}^\infty L_{m}\sum_{n=0}^{m-1} G_n(\lambda)\nu^{m-n-1}
\label{defR}
\ee
and the contour $C_2$ encircles both the cut singularity of
$G_{\om\l}$ and the pole at $\om=\nu$.  The terms on the
right-hand-side result from taking the residue at infinity in the
contour integral and the functions $G_n(\l)$ are defined by \eq{bcG}.

The formal solution to \eq{gen eq for G} is
\be
G_{\nu\l} = \frac{E_\nu-R_\l(\nu)}{\l+E_\nu- L(\nu)}
\label{genG}
\ee
with $R_\l(\nu)$ given by~\rf{defR}.%
\footnote{It is worth mentioning that \rf{genG} obviously satisfies
\eq{Migdal} as is discussed in the previous section since $R_\l(\nu)$
has no discontinuity in $\nu$ at the cut.} The functions $G_n(\l)$ can
be expressed in terms of $E_\l$ and $L(\omega)$ using the recurrence
relation
\be
G_{n+1}(\l)=\ci \frac{L(\om)}{\l-\om} G_n(\om) - E_\l G_n(\l)~,
{}~~~~~~G_0(\l)=E_\l
\label{recurrent}
\ee
which can easily be obtained by expanding \eq{main} in $1/\l$.  For
$n=0$ this equation recovers \eq{1}.

It still remains to determine $E_\l$. It can be determined from the
equation
\footnote{The equation which was solved in Ref.~\cite{Mak93} by the
ansatz $F(\om)=\om^2 L^{-1}(\om)$ is just the $1/\l^2$-term of the
expansion of \eq{smart integral} in $1/\l$ and is, therefore, incomplete.
This ansatz is not consistent, generally speaking, with the whole
\eq{smart integral}.}
\be
\ci \, L(\om) G_{\om\l} =  \l E_\l -1
\label{smart integral}
\ee
which is the $1/\nu$ term of the asymptotic expansion of \eq{main} in
$\nu$. Using the expansion~\rf{bcG}, one can rewrite this equation as
\be
\sum_{m\geq0} L_{m}G_{m}(\l) = \l E_\l -1~.
\label{smart}
\ee

If $L(\l)$ were a polynomial of the highest power $J$,
\eq{smart} and \eq{recurrent} yield a polynomial equation  for
$E_\l$ containing powers of $E_\lambda$ up to order $J$.
Some explicit solutions of this equation are given in the next
subsections.  Once
$E_\lambda$ is known, $\tilde V$ can be obtained from \eq{tildeV}.

One could consider also the next terms of the $1/\nu$-expansion of
\eq{main} which result in the equations
\be
\sum_{m\geq0} L_{m}G_{m+n}(\l) = \l G_n(\l)
+ \sum_{k=0}^{n-1} E_k G_{n-1-k}(\l)  - E_n~,~~~~~~~n\geq1~.
\label{automatic}
\ee
A question arises as to whether these equations impose new
restrictions on $E_\l$. The answer is ``no'' and \eq{automatic} is
automatically satisfied as a consequence of Eqs.~\rf{recurrent} and
\rf{smart}.  The proof can be given by induction. Let us calculate the
left-hand-side of \eq{automatic} at some $n=n_0$ assuming that it
holds for all lower $n<n_0$. One gets
\bea
{}& \sum_{m\geq0} L_{m}G_{m+n}(\l)
\begin{array}[b]{c}\hbox{\footnotesize (\ref{recurrent})} \\ = \end{array}
\sum_{m\geq0} L_{m}\left\lbrace \ci \frac{ L(\om)}{\l-\om} G_{m+n-1}(\om)
- E_\l G_{m+n-1}(\l)\right\rbrace
\begin{array}[b]{c}\hbox{\footnotesize (\ref{smart})} \\ = \end{array} & {}
\non {} &
\l \left\lbrace \ci \frac{ L(\om)}{\l-\om} G_{n-1}(\om) - E_\l
G_{n-1}(\l) \right\rbrace +
\sum_{k=0}^{n-1} E_k G_{n-1-k}(\l)
- \ci  L(\om) G_{n-1}(\om)& {} \non {} &
\begin{array}[b]{c}\hbox{\footnotesize (\ref{recurrent})} \\ = \end{array}
\l G_n(\l) + \sum_{k=0}^{n-1} E_k G_{n-1-k}(\l)  - E_n &{}
\eea
which completes the proof.

As is discussed above, $G_{\nu\l}$ should be symmetric in $\nu$ and $\l$.
While the right-hand-side of \eq{genG} with $R_\l(\nu)$ given by
\eq{defR} looks non-symmetric, it can be shown that the symmetry is
restored as a consequence of \eq{smart}. To prove this, let us multiply
the numerator and denominator on the right-hand-side of \eq{genG} by
$(\nu+E_\l- L(\l))$. The denominator then becomes explicitly symmetric
and the numerator can be transformed as
\bea
& \Big(E_\nu-R_\l(\nu)\Big)\Big(\nu+E_\l- L(\l)\Big)
\begin{array}[b]{c}\hbox{\footnotesize (\ref{recurrent})} \\ = \end{array}
 E_\nu E_\l + \nu E_\nu -E_\nu L(\l)&\non &
+ \sum_{m=1}^\infty L_{m} \sum_{n=0}^{m-1}
\nu^{m-n-1}
\left\lbrace G_{n+1}(\l)
-\cii \frac{ L(\om)}{\l-\om} G_{n}(\om) - \nu
G_{n}(\l) \right\rbrace
\begin{array}[b]{c}\hbox{\footnotesize (\ref{defR})} \\ = \end{array}& \non
&
E_\nu E_\l + \nu E_\nu - E_\nu L(\l) - E_\l L(\nu)
+ \sum_{m=0}^\infty L_{m} G_{m}(\l)
+ \cii \frac{L(\om)}{\l-\om}\int_{C_2}\frac{dz}{2\pi i}
\frac{ L(z)}{\nu-z} G_{\om z} &
\non
& \begin{array}[b]{c}\hbox{\footnotesize (\ref{smart})} \\ = \end{array}
E_\nu E_\l - E_\nu L(\l) - E_\l L(\nu) + \nu E_\nu +\l E_\l -1
+ \cii \frac{ L(\om)}{\l-\om}\int_{C_2}\frac{dz}{2\pi i}
\frac{ L(z)}{\nu-z} G_{\om z}~.&
\eea
This expression is manifestly symmetric in $\nu$ and $\l$, what is easy to see
taking the imaginary part, which completes
the proof.

As is shown
below for the quadratic and quartic cases,
the symmetry requirement can be used directly to determine $E_\l$
alternatively to \eq{smart}.

\subsection{Quadratic potential}

Let us show how the algorithm of the previous subsection works for the
simplest case of a Gaussian potential.  In this case, we begin with
the assumption that $L_{m}=0$ for $m\neq1$ so that $L(\l)$ is
\be
L(\l) = L_1 \l \equiv\L^{-1} \l\,,
\ee
where we have introduced the quantity $\L$ to establish the connection
with the previous studied of the Gaussian model.

In the Gaussian case the solution~\rf{genG} takes the form
\be
G_{\nu\l}^{(0)} = \frac{E_\nu-\La^{-1}E_\lambda}{\l+E_\nu-\nu \La^{-1}}
\label{GG}
\ee
since $R_\l(\nu)=\L^{-1}E_\l$. Requiring this expression to be symmetric w.r.t.
$\la$ and $\nu$ one fixes the function $E_\la$ completely. It turns out that
$E_\l$ coincides with the function obtained from the
semi-circle eigenvalue distribution of the Gaussian matrix model
(which is also known to solve the Gaussian Kazakov-Migdal model
\cite{Gro92}),
\be
E_\l^{(0)} = \frac{\mu\l}{2}-\sqrt{\frac{\mu^2\l^2}{4}-\mu}~,
{}~~~~\rho^{(0)}(x) = {1\over\pi} \sqrt{\mu-\frac{\mu^2\l^2}{4}}~~,
\label{Gauss E}
\ee
while $\mu$ and $\L$ are related by
\be
\mu=\La^{-1}-\La
\label{Gaussmu1}.
\ee
Therefore, the  symmetry requirement  fully determines the Gaussian
solution.

This allows to rewrite $G_{\nu\l}^{(0)}$  in the
manifestly symmetric form
\be
G_{\nu\l}^{(0)}=\frac{\L^{-1}E_\l E_\nu-\L^{-1} (\l E_\l+\nu E_\nu)+
(\l E_\nu+\nu E_\l)+\mu} {\l^2-\left( \L+\L^{-1} \right) \l\nu +
\nu^2+\mu} ~.
\label{GaussianG}
\ee
It is straightforward to check that this expression is not singular for all
$\nu$ and $\la$.

As for the equation~\rf{smart}, in the Gaussian case it involves only
$G_1(\l)$ which is fixed by the recurrence relation~\rf{recurrent} to
be
\be
G_1^{(0)}(\l)= \frac{1}{\L} (\l E_\l-1) -E_\l^2~.
\ee
This gives the quadratic equation
\be
{1\over\La}E_\l^2+\left(1-{1\over\La^2}\right)\l E_\l = 1-{1\over\La^2}
\ee
whose solution, which at large $\l$ behaves like $1/\l$, is
given by Eqs.~\rf{Gauss E}, \rf{Gaussmu1}.

Since (\eq{defG1})
\be
G_1(\nu) = \int_{C_1} \frac{d\om}{2\pi i} \frac{F(\om)}{\nu-\om}E_\om =
\int^b_a dx \rho(x) \frac{F(x)}{\nu-x}
\ee
where the last integral is along the cut, $F(x)$ is easily found from the
discontinuity of $G_1(x)$ across the cut. The result coincides with that of
\cite{Mak92}
\be
F^{(0)}(x) = \La x
=\frac{2x}{\mu+\sqrt{\mu^2+4}} \,.
\label{GaussLambda1}
\ee
It is crucial that $D$ does not enter explicitly
in the relation between $\mu$ and $\L$
(cf. Eqs.~\rf{Gaussmu1}, \rf{GaussianG}). \sloppy
The meaning is that the Kazakov--Migdal model
with the quadratic potential reduces to
the Hermitean one-matrix model with the same kind of potential.

Taking discontinuities of \rf{GaussianG} across the cuts according to the
definition \rf{defG} we obtain the following expression for the correlator of
the gauge fields
\be
C(\a,\b)=\frac{\L^{-1}}{\a^2-\left( \L+{\L}^{-1} \right) \a\b
+\b^2+\mu}\,.
\label{GaussianC}
\ee

Its expansion as $\mu\ra\infty$ recovers the leading order of the large mass
expansion~\cite{KMSW92a}
\be
C_{ij}=1+\phi_i\phi_j \,.
\ee
On the contrary if $\mu\ra0$, one gets from~\rf{GaussianC} the weak
coupling result~\cite{DKSW93}
\be
C_{ij}=N \delta_{ij} \,.
\ee
While the expression~\rf{GaussianC}
involves the quadratic polynomial in the denominator,
its zeros, say in the complex $\l$-plane,
\be
\l_\pm(\nu)~=~\frac 12 \left( \L+\frac 1\L \right) \nu \pm
\frac \mu2\sqrt{\nu^2-\frac 4\mu}\,,
\label{GaussZeros}
\ee
always lies for $\mu>0$ outside the cut.

The situation when $C(\a,\b)$ were become negative somewhere at
the support of $\rho$, would be associated with a (third order)
phase transition in the Itzykson--Zuber integral. The only possibility
to have such a phase transition with our Gaussian \eq{GaussianC} is when
$\mu=0$ so that $\l_\pm=\nu$ belongs to the cut. Because~\cite{Gro92}
\be
m_0^2 = D\sqrt{\mu^2+4} -(D-1)\mu \,,
\label{m}
\ee
the only possibility to reach the point $\mu=0$ in the strong coupling phase
$m_0^2\geq2D$, where the Gaussian model is stable, is at $D=1$. This
phase transition at $D=1$ is the standard one which provides
the continuum limit.

\subsection{The quartic case}

One of simplest non-Gaussian cases is associated with
\be
L(\l)=L_1 \l + L_3 \l^3~.
\label{ansatz}
\ee
The meaning of this formula is that we consider \eq{ansatz} as an {\it
ansatz}\/ for $L(\l)$ and  shall solve~\eq{main} for $E_\l$ (and,
therefore, $\tilde{V}'(\l)=2\real E_\l$ at the cut). The original potential
$V$ of the Kazakov--Migdal model can then be determined due to \eq{defL}
to be
\be
V^\p(\l)= 2D  L(\l) - (2D-1) \tilde{V}^\p(\l)~.
\ee
As was already discussed, $D=1/2$ for the two-matrix model and
$\tilde{V}^\p$ disappears from this formula. For this reason one might
think of $\int L(\l)d\l$ as of the potential of the proper two-matrix model
which is just quartic for the ansatz~\rf{ansatz}.

For $L(\l)$ given by \eq{ansatz} only the term with $m=3$ survives
in~\rf{defR} so that \eq{genG} gives
\be
G_{\nu \la} = \frac{E_\nu - (L_1+L_3\nu^2)E_\la -L_3\, \nu\, G_1(\la)-
 L_3 G_2(\la)}{\l + E_\nu- L(\nu) }
\label{G}
\ee
where the recurrence relation~\rf{recurrent} yields
\bea
&G_1(\la) = (L_1+L_3\la^2) (\la E_\la-1) - E^2_\la - L_3 E_2
\non
&G_2(\la) = \left(  L(\la) - E_\la \right) G_1(\la) -
\l L_3 (L_1 E_2 + L_3 E_4 -1)
\label{G12}
\eea
with $E_2$ and $E_4$ defined by \eq{bc1}.

To determine $E_\l$ we use \eq{smart} (or, which is equivalent, the
requirement that $G_{\nu \la}$ should be symmetric in $\nu$ and $\la$)
which leads to the following {\it quartic}\/ equation
\be
 E_\l \left( \l-L_1 L(\l)\right) +L_3 G_2(\l)
\left( E_\l- L(\l)\right) +
L_1 E^2_\l+L_3 \l ^2(L_1+L_3E_2) = {\rm const}~.
\label{E}
\ee
The constant on the right-hand-side of this equation can be found through
asymptotic expansion in $\l$ of its left-hand-side.

Since the resulting equation \re{E} for $E_\nu$ is quartic, its general
solution,
 though explicitly known, is rather obscure and not informative.
For our purposes it is more convenient to rewrite it lowering
 the powers of $E_\l$  down
to linear using \eq{EE} which we rewrite in the form
\be
E_\l^2 = \tilde{V}^\p(\l) E_\l -Q(\l)
\label{EEE}
\ee
where
\be
Q(\l) \equiv - \cii \frac{\tilde{V}^\p(\om)}{\l-\om} E_\om
\label{defQ}
\ee
is a polynomial of the highest power $J-2$ if is $\tilde{V}$ is
that of the highest power $J$. It is essential for what follows that
nether $Q(\l)$ nor $\tilde{V}^\p(\l)$ are singular at the cut of $E_\l$.

The resulting equation can be conveniently rewritten in terms of
$F(\l)=L(\l)-\tilde{V}^\p(\l)$.
Taking the discontinuity of the resulting equation (\ie just the factor
in front of $E_\l$) one gets
\bea
& & L_1F(\l) + L_3\Big[
F^3(\l) -F(\l)Q(\l)- \Big(Q(\l)-(L_1+L_3\l^2)-
 L_3E_2\Big)\Big(F(\l)+L(\l)\Big)  \non
& & + \l L_3 (L_1E_2+L_3E_4-1)\Big] =\l
\label{sym1}
\eea
while the continuous part yields
\bea
& L_1Q(\l)+L_3\Big[F(\l)Q(\l)\Big(F(\l)+L(\l)\Big) & \non & -
\Big(Q(\l)-(L_1+L_3\l^2)-L_3E_2\Big)\Big(Q(\l)-L^2(\l)\Big)
&\non & - L_3\l L(\l) (L_1E_2+L_3E_4-1)\Big] =L_3\l(L(\l)+\l
L_2E_2)+~\hbox{const.}&
\label{sym2}
\eea

The set of equations~\rf{sym1}, and \rf{sym2} must have a solution
with $F(\l)$ and $Q(\l)$ nonsingular at the cut.  In general, $Q(\l)$
should be expressed via $F(\l)$, say, from
\eq{sym1}. The substitution into \eq{sym2} then yield a high order
algebraic equation for $F(\l)$. Presumably, a reasonable way to study
the equation for $F(\l)$ is to assume some analytic structure and then
determine algebraic coefficients from the equation. From a bunch of
solutions one should choose the one which recovers the Gaussian
solution as $L_3\ra 0$.

While the general solution of Eqs.~\rf{sym1}, \rf{sym2} looks
hopeless, the iterative solution in $L_3$ is straightforward
\be
F(\l)=\left(\frac{1}{L_1}-\frac{2L_3}{L_1^3}\right) \l
-\frac{L_3}{L_1^4}\l^3
\ee
and
\be
\tilde{V}^\p(\l)~=~\left( L_1-\frac{1}{L_1}+\frac{2L_3}{L_1^3}\right)\l
+L_3\left(1+\frac{1}{L_1^4}\right)\l^3\,.
\ee
Next orders in $L_3$ can easily be calculated which illustrates that
the solution exists at least at small enough $L_3$.

Given the solution to Eqs.~\rf{sym1}, \rf{sym2}, $C_{ij}$ can be
calculated taking the double discontinuity of~\rf{G} w.r.t.\ both
$\nu$ and $\l$ across the cut which gives
\be
C(x,y)=\frac{(L_1+L_3x^2)+L_3xF(y)+L_3\Big[F^2(y)-Q(y)+
(L_1+L_3y^2)+L_3E_2 \Big]} {y^2-y\Big( L(x)+F(x) \Big)+L(x)F(x)+Q(x)}
\,.
\label{Cquartic}
\ee
As is discussed, this expression will be symmetric in $x$ and $y$ for
$F$ and $Q$ being the solutions of Eqs.~\rf{sym1}, \rf{sym2}.  This
requirement of the symmetry of~\rf{Cquartic} might help to find a
solution. Anyway, the ansatz~\rf{ansatz} was choosen as the simplest
one while for other functions $L(\l)$ the resulting algebraic
equations might simplify.

\newsection{Discussion}

In this Paper we have developed a method for calculating the gauge
field correlator $C(\a,\b)$. Its knowledge allows one to calculate the
{\it extended\/} loop averages
\be
G_{\nu\l}(\Gamma_{xy}) \equiv \LA \ntr \Big( \frac{1}{\nu-\phi_x}
U(\Gamma_{xy})\frac{1}{\lambda-\phi_x} U(\Gamma_{yx})\Big) \RA
\label{extended}
\ee
where $U(\Gamma_{xy})$ is the ordered product of $U$'s along some path
$\Gamma_{xy}$ from $x$ to $y$. If $\Gamma_{xy}$ coincides with one
link, $G_{\nu\l}(\Gamma_{xy})$ coincides with $G_{\nu\l}$ according to
the definition~\rf{defG}. An explicit formula for
$G_{\nu\l}(\Gamma_{xy})$ can be obtained substituting the gauge field
correlator at each link $l$ by $C(\a_l,\a_{l+1})$ and integrating over
$\a_l$'s.

In the case of a matrix chain which is associated with the
Kazakov--Migdal model at $D<1$, this would be a complete set of
observables at $N=\infty$. However, a new class of correlators which can be
constructed from pure gauge fields appears at $D>1$. The simplest one is
the closed {\it adjoint} Wilson loop
\be
W_A \equiv \LA \left| \ntr U(\Gamma) \right|^2 \RA
\label{adjoint}
\ee
which is well-defined in the 't~Hooft limit~\cite{lgn} and can again
be calculated integrating over $U$'s at each link. It is
expected~\cite{KSW92,KhM92} that this quantity might undergo a
first-order phase transition which is associated with a restoration of
the area law. For the Gaussian model this phase transition does not
happen while more complicated potentials have not yet been studied.

One more interesting problem, for which the approach of this paper
might be potentially useful, concerns another kind of pure gauge field
averages, the {\it filled\/} Wilson loops which were employed in
connection with the Kazakov--Migdal model in~\cite{KSW92} and whose
importance for constructing the $D>1$-dimensional string theories from
the Kazakov--Migdal model was emphasized in~\cite{KMSW92a} and, as
well as their role in resolving the problem of the local $Z_N$
symmetry, in ~\cite{DKSW93}.

We hope that the technique developed in this paper could help to solve
some of these problems.

\section*{Acknowledgements}
We are grateful to J.~Ambjorn, C.~Kristjansen and N.~Weiss for useful
discussions. Yu.~M. thanks the UBC physics department for the
hospitality at Vancouver.  This work is supported in part by the
Natural Sciences and Engineering Research Council of Canada.

%\eop


\begin{thebibliography}{31}
\small
\addtolength{\itemsep}{-6pt}

\bibitem{lgn}G. 't Hooft, {\sl Nucl. Phys.} {\bf B72} (1974) 461.

\bibitem{KM92}
V.A. Kazakov and A.A. Migdal,
%{\it Induced QCD at large $N$},
{\sl Nucl.~Phys.} {\bf B397} (1993) 214.

\bibitem{itzub}
C. Itzykson and J.B. Zuber, {\sl J. Math. Phys.} {\bf 21} (1980) 411; \\
%\bibitem{harish}
Harish-Chandra, {\sl Amer. J. Math.} {\bf 79} (1957) 87.

\bibitem{Mig92a}
A.A. Migdal,
%{\it Exact solution of induced lattice gauge theory at large $N$},
{\sl Mod.~Phys.~Lett.} {\bf A8} (1993) 359.

\bibitem{Gro92}
D. Gross,
%{\it Some remarks about induced QCD},
{\sl Phys.~Lett.} {\bf 293B} (1992) 181.

\bibitem{Mak92}
Yu.\ Makeenko,
%{\it Large-N reduction, master field and loop equations in
%the Kazakov--Migdal model},
{\sl Mod.~Phys.~Lett.} {\bf A8} (1993) 209.

\bibitem{KMSW92a}
I.I. Kogan, A. Morozov, G.W. Semenoff and N. Weiss,
%{\it Area law and continuum limit in "induced QCD"},
{\sl Nucl. Phys.} {\bf B395} (1993) 547.

\bibitem{DKSW93}
M.I.~Dobroliubov, I.I.~Kogan, G.W.~Semenoff and N.~Weiss,
{\sl Phys.~Lett.} {\bf 302B} (1993) 283.

\bibitem{KSW92}
I.I. Kogan, G.W. Semenoff and N. Weiss,
%{\it Induced QCD and hidden local $Z_N$ symmetry},
{\sl  Phys. Rev. Lett.} {\bf 69} (1992) 3435.

\bibitem{KhM92}
S. Khokhlachev and Yu.~Makeenko,
%{\it The problem of large-N phase
%transition in Kazakov-Migdal model of induced QCD },
{\sl Phys. Lett.} {\bf 297B} (1992) 345.

\bibitem{MorShat}
A.~Morozov,
%{\it Pair correlator in the Itzykson-Zuber integral},
{\sl Mod.~Phys.~Lett.} {\bf A7} (1992) 3503; \\
S.~Shatashvili, {\it Correlation functions in the Itzykson-Zuber model},
preprint IASSNS-HEP-92/61 (September, 1992).

\bibitem{Sta93}
M.~Staudacher, {\sl Phys.~Lett.} {\bf 305B} (1993) 332.

\bibitem{AK}
J.~Ambjorn and C.~Kristjansen, unpublished.

\bibitem{Mak93}
Yu.~Makeenko, {\it An exact solution of induced large-$N$ lattice
gauge theory at strong  coupling}, preprint ITEP-YM-2-93 (March, 1993).


\end{thebibliography}
\end{document}